\documentclass[runningheads]{llncs}

\usepackage{xcolor}
\usepackage{makecell}
\usepackage{multicol}
\usepackage{multirow}
\usepackage{adjustbox}
\usepackage{mdwtab}
\usepackage{anyfontsize}
\usepackage{times}
\usepackage{epsfig}
\usepackage{amsmath}
\usepackage{amssymb}
\usepackage{tocloft}
\usepackage{gensymb}
\usepackage{hyperref}
\hypersetup{
    colorlinks=true,
    linkcolor=blue,
    filecolor=magenta,      
    urlcolor=red,
}
\usepackage{graphicx}
\begin{document}
\title{Edge Profile Super Resolution}
%
\author{Jiun LEE\inst{1}\orcidID{0000-0002-7338-4565} \and
 Inyong YUN\inst{2}\orcidID{0000-0001-8082-033X}\and
Jaekwang KIM\inst{1}\orcidID{0000-0001-5174-0074}}
\authorrunning{J. LEE et al.}
%
\institute{Sungkyunkwan University, Suwon 16419, Republic of Korea\\
\email{\{jiwoon94, linux\}@skku.edu}\\\and
Big Data \& AI Lab, Hana Institute of Technology, Hana TI, Seoul 06133, Republic of Korea\\
\email{iyyun@hanafn.com}}
\maketitle              
\begin{abstract}
In this paper, we propose Edge Profile Super Resolution(EPSR) method to preserve structure information and to restore texture. We make EPSR by stacking modified Fractal Residual Network(mFRN) structures hierarchically and repeatedly. mFRN is made up of lots of Residual Edge Profile Blocks(REPBs) consisting of three different modules such as Residual Efficient Channel Attention Block(RECAB) module, Edge Profile(EP) module, and Context Network(CN) module. RECAB produces more informative features with high frequency components. From the feature, EP module produce structure informed features by generating edge profile itself. Finally, CN module captures details by exploiting high frequency information such as texture and structure with proper sharpness. As repeating the procedure in mFRN structure, our EPSR could extract high-fidelity features and thus it prevents texture loss and preserves structure with appropriate sharpness. Experimental results present that our EPSR achieves competitive performance against state-of-the-art methods in PSNR and SSIM evaluation metrics as well as visual results.
\end{abstract}
\section{Introduction}
Single Image Super-Resolution(SISR)\cite{freeman2000learning} has been focused on recently. Generally, SISR targets to reconstruct an accurate high resolution(HR) image from its degraded low resolution(LR) image. Image super-resolution(SR) is usually applied to diverse computer vision tasks (e.g. security and surveillance imaging\cite{zou2011very}, object recognition\cite{sajjadi2017enhancenet}, image generation\cite{karras2017progressive}, and medical imaging\cite{shi2013cardiac}). Since there are plenty of solutions for reconstructing any LR inputs, image SR has an ill-posed inverse\cite{ebrahimi2007solving} problem. For high-fidelity image, it is necessary to represent details including high frequency components such as texture and structural information. To address this issue, numerous SR methods have been proposed, such as conventional methods\cite{zhang2006edge,dong2011image,fattal2007image,sun2010gradient,yan2015single} and deep learning methods\cite{zhang2018residual,lim2017enhanced,ledig2017photo,lai2017deep,wang2018esrgan,soh2019natural}.\\
\indent In conventional methods, edge-based models\cite{fattal2007image,sun2010gradient,yan2015single} enhance sharpness of super resolved image by utilizing edge statistics. They model edge statistical dependencies by estimating structural connectivity between HR and LR. However, edge distribution tends to be heavily dependent on the similarity between training and test datasets. Therefore, the performance lacks consistency. Furthermore, since they focus on sharpness of SR image, they have weakness on improvement of texture restoration. The modeling is proceeded point by point. Hence, the process of edge generation is complex and inflexible.\\ 
\indent On the other hand, deep learning methods are more flexible and remarkable in handling probability transformation including pixel distribution. They acquire outstanding results compared with previous methods\cite{he2016deep,chang2004super} recently. Normally, deep learning methods approach SISR problem by utilizing influential feature representation and deep end-to-end structure. These models\cite{lim2017enhanced,lai2017deep,zhang2018residual} achieve notable improvement in visual quality. In this case, most of them are optimized as measuring pixel distance between SR and its corresponding HR by MSE or $L_1$. This optimizing methods tend to make the networks generate an image based on statistical information of possible HR solutions. Even though they reach high numerical value evaluation on peak signal-to-noise ratio(PSNR), general deep learning models show blurry with texture loss and structural trouble results.\\
\indent To represent texture and preserve image structure, Yang et al. \cite{yang2017deep} applies simply edge information in deep learning model. However, they utilize edge information as assistant device and design their model to reach higher PSNR evaluation metric and thus using structure information is inadequate. For perceptual improvement, some methods such as \cite{ledig2017photo,wang2018esrgan,soh2019natural} utilize the generative adversarial network(GAN) with perceptual loss to generate photo realistic image. 
 Although these perceptual-driven models bring perceptual enhancement by restoring texture information related to blurry problems, they can not avoid structural distortions in details with definite edges. To overcome the image structural limitation, some models\cite{ma2020structure,nazeri2019edge} utilize structural information by designing additional module for preserving structure. The models feed explicit guidance to an established perceptual-driven model for solving structure problems in SR. Even though they compensate the structural defects of GAN-based model, they do not still reach the visual quality of HR images. Furthermore, since the discriminators may bring unstable factors during optimization procedure, GAN-based models have difficulties in stability of learning process and keeping structural consistency. \\
\indent In this paper, we propose an Edge Profile Super Resolution(EPSR) method to alleviate the issues that we have mentioned above. In SISR problem, to generate high quality SR image, it is important to represent high frequency details such as structure and texture information. Since these components have frequent pixel variations, they have contextual properties and thus displaying them is the crucial point for high quality results. To achieve the goal, we modify Fractal Residual Network(FRN) as network structure to utilize various information in learning process. we call it modified Fractal Residual Network(mFRN) structure. To draw high frequency components from diverse information, we construct Residual Edge Profile Blocks(REPBs) as basic blocks. REPB consists of Residual Efficient Channel Attention Block(RECAB) module, Edge Profile(EP) module and Context Network(CN) module. For extracting high-fidelity features, it is necessary to utilize informative features which contain detail information. Hence, by referring to previous methods\cite{lim2017enhanced,zhang2018image} and recent research\cite{qilong2020eca}, we apply ECA on feature extraction. This systemically organized feature provide abundant information to EP module. EP module feeds thus structural information on the features by generating edge profile itself from the informative features. This module is based on principle of conventional edge extraction. Even though EP module contributes to preserving image structure, exploiting high frequency components such as sharpness and textures should be considered for more high-fidelity results. However, these contextual details contain complex variations in specific regions(i.e. high frequency regions such as edge and texture) and thus there are difficulties to maintain the detail information in process.
 To exploit high frequency components, we construct Context Network(CN) module. By exposing contextual information, this module captures pixel variation and thus sharpness of results could be enhanced properly and texture loss also be restored.
 By proceeding repeatedly this process in network, SR results shows structural stability and representing details with reducing texture loss and structure distortions. Experimental results on benchmark datasets demonstrate that our EPSR achieves in improving SR quality.
\section{Related Work}
In the computer vision community, various SISR methods have been proposed for several years. To be related with our proposed method, we review on SISR methods into three categories: Edge-related methods\cite{fattal2007image,sun2010gradient,yan2015single,tai2010super,zhu2015modeling,yang2017deep,ma2020structure,nazeri2019edge}, General deep learning method\cite{dong2015image,kim2016accurate,kim2016deeply,tai2017image,tai2017memnet,lim2017enhanced,zhang2018residual,zhang2018image,dai2019second} and Perceptual-driven method\cite{johnson2016perceptual,ledig2017photo,sajjadi2017enhancenet,wang2018esrgan,wang2018recovering,rad2019srobb}\\
\textbf{General deep learning method}\\
Recently, general deep learning methods have been mainly studied in single image super resolution. SRCNN proposed by Dong et al.\cite{dong2015image} achieves noteworthy performance using three-layer convolutional network. Later, VDSR\cite{kim2016accurate} and DRCN\cite{kim2016deeply} improve accuracy with stacking convolutional networks deeply through residual learning. Tai et al.\cite{tai2017image} introduce DRRN, which is a recursive learning model based on parameters sharing and they propose MemNet\cite{tai2017memnet}, which consists of memory block for a deep network. EDSR and MDSR by Lim et al.\cite{lim2017enhanced} improve significantly the performance by stacking residual blocks very deeply and widely. From the results, the depth of network is a key point in image SR. Since the achievement of deep networks, RDN by Zhang et al.\cite{zhang2018residual} is designed as a deep network based on the dense block for utilizing all of the hierarchical features from all the convolutional layers. Zhang et al.\cite{zhang2018image} and Dai et al.\cite{dai2019second} consider not only increasing the depth of network, but also applying feature correlations in spatial and channel dimension. From the investigations, general deep learning methods target to achieve high PSNR performance by utilizing feature information efficiently.\\
\textbf{Perceptual-driven method}\\
As aforementioned, all general deep learning methods concentrate on achieving high PSNR. However, their results display blurry and unstable structural SR images. For recovering SR image more toward realistic direction, Johnson et al.\cite{johnson2016perceptual} propose perceptual loss to enhance the visual quality of SR images. Ledig et al.\cite{ledig2017photo} design SRGAN based on adversarial loss and it is the first model that can generate  photo-realistic HR images. EnhanceNet by Sajadi et al.\cite{sajjadi2017enhancenet} shows high-fidelity textures SR images by applying texture loss. Wang et al.\cite{wang2018esrgan} propose ESRGAN which enhances the previous frameworks by constructing Residual-in Residual Dense Block(RRDB). On the other hand, Wang et al.\cite{wang2018recovering} generate more natural textures for specific categories by exploiting semantic segmentation maps as priors. In addition, SROBB by Rad et al.\cite{rad2019srobb} is proposed to a objective perceptual loss based on the labels of object, background and boundary. These perceptual-driven methods shows enhancement in overall visual quality. However, they leave problems of structure distortions and fails recovering details such as texture. \\
\textbf{Edge-related method}\\
Edge and gradient information has been utilized in previous SISR works. Fattal \cite{fattal2007image} proposes a method learning the prior dependencies among edge statistics of image gradients. Sun et al. \cite{sun2010gradient} propose a gradient field transformation to control HR gradient fields and enhance sharpness. Yan et al.\cite{yan2015single} propose a method based on gradient profile sharpness extracted from gradient description models. Tai et al.\cite{tai2010super} propose an approach to combine edge-directed SR with detail from an image and texture examples. These models are dependent on connectivity and relation HR and LR. Thus the results are decided by similarities between train and test datasets. Furthermore, since the processes are modeled point by point, they are complicated and less flexible. Zhu et al.\cite{zhu2015modeling} propose a SISR method based on the gradient reconstruction by collecting a dictionary of gradient patterns. Yang et al.\cite{yang2017deep} propose a recurrent residual network which applies edge information from off-the shelf edge detector. However, this method targets to restore high-frequency components related to PSNR evaluation. Ma et al.\cite{ma2020structure} utilizes edge information in perceptual-driven methods as explicit guidance and Nazari et al.\cite{nazeri2019edge} propose an edge-informed SR method based on image inpainting task. They contribute to preserving structural information. However, they still have weakness in recovering high frequency components such as sharpness and texture. For high-fidelity image, it is important to represent high frequency components. To draw high quality results, our proposed method aims to exploit high frequency information that are related to structure with proper sharpness and texture by utilizing visual and their contextual properties.

\section{Methodology}
\indent In this section, we present the overview of the EPSR. Then we introduce the details of REPB which forms informative features by utilizing structural information and exploiting high frequency components. At the end, we describe objective functions. 
\subsection{Overview}
\begin{figure*}[h]
\begin{center}
\includegraphics[width=11.0cm]{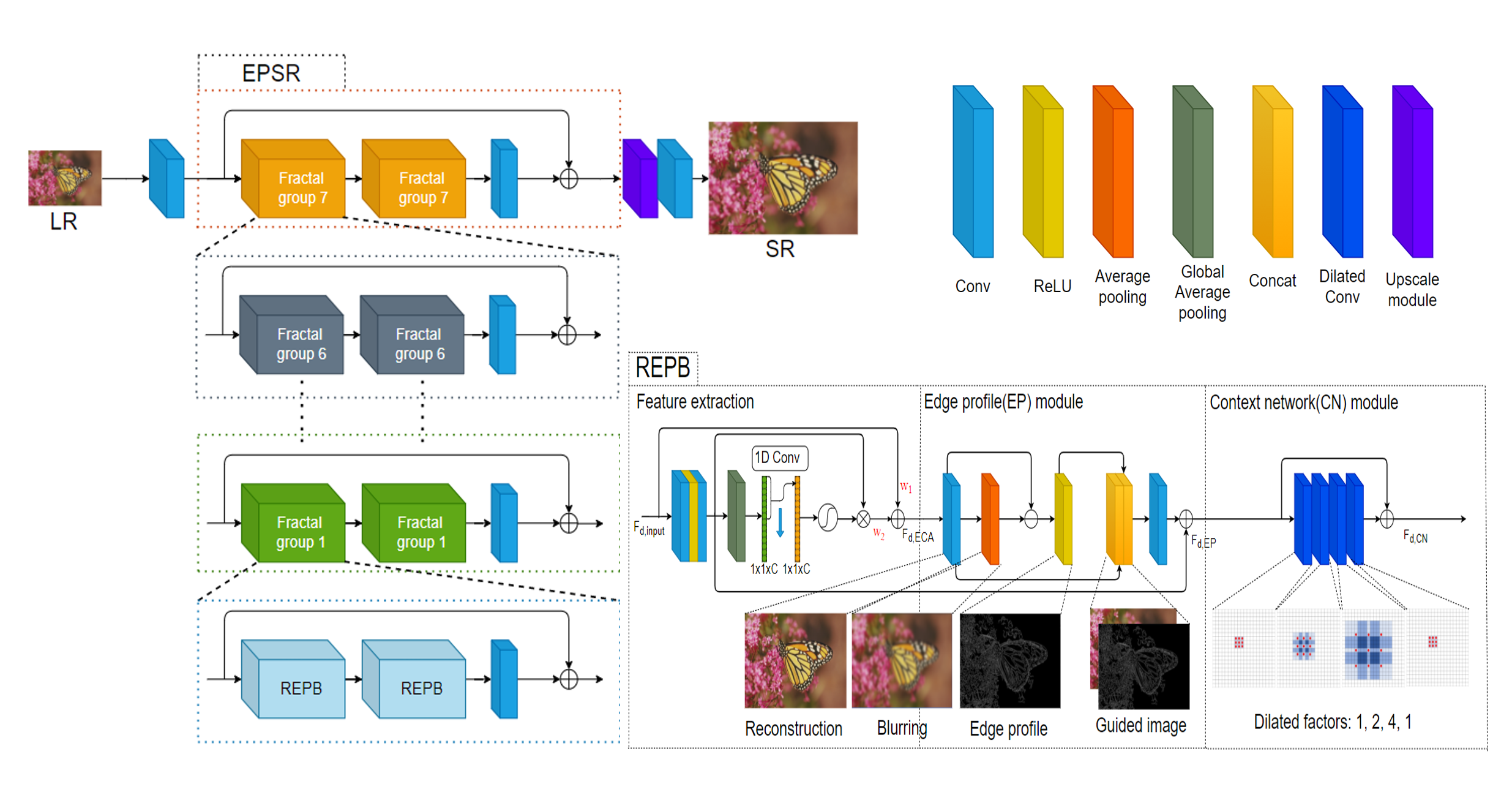}
\end{center}
\caption{The architecture of the proposed EPSR which follows mFRN structure consisting of REPBs as basic blocks. REPB consists of RECAB module, EP module and CN module. we also include the explanation of components of all blocks.}
\label{Fig.2}
\end{figure*}
The overall structure is described in Fig.\ref{Fig.2}. As researched in \cite{lai2018fast,zhang2018residual}, we apply one convolution layer to extract the shallow feature from the LR input. To utilize various information in process, we modify the skip connection structure of FRN by Kwak et al.\cite{kwak2019fractal} and modify as in Fig.\ref{Fig.2}. we call it mFRN. mFRN consists of REPBs. Since the self-similarity property of mFRN structure gains deep depth and provide very large receptive field size, REPBs can obtain diverse information and generate informative features effectively, which include high frequency components containing details such as structure with sharpness and texture. Then the deep features from the mFRN structure is upscaled by upscale module. We apply this upscaling module such as previous work \cite{dong2016accelerating,zhang2018residual}. According to the process, the upscaled feature is then converted into SR image via one convolution layer.
\subsection{Residual edge profile block(REPB)}
Due to self-similarity of mFRN structure, the abundant diverse frequency information can be bypassed. From various information, Our proposed REPBs can focus on exploiting high frequency components by utilizing influential features with structural information and exposing contextual information. Our REPB consists of three parts: RECAB module, EP module and CN module. \\
\textbf{Residual Efficient Channel Attention Block(RECAB)}:
As proposed in EDSR, MDSR \cite{lim2017enhanced}, by removing batch normalization layers, we extract the feature. Thus range flexibility of our EPSR can be maintained. So we can formulate feature extraction as
\begin{equation}
F_{FE}= H_{FE}(F_{input}),
\end{equation}
where the output $F_{FE}$ and $H_{FE}(\cdot)$ stand for the feature and function from feature extraction of REPB block respectively. $ F_{input}$ is the input feature of REPB block. In SISR problem, RCAN by Zhang et al.\cite{zhang2018image} consider feature interdependencies and utilizing mutual independence by applying channel attention process from SENet\cite{hu2018squeeze}. However, this process has been shown that dimensionality reduction brings side effects on channel prediction. By messing up the direct correspondence between its channel and weight, it captures unnecessary dependencies across all channels empirically. To avoid this problem, we use efficient channel attention(ECA) by \cite{qilong2020eca}. ECA captures local cross-channel interaction by using 1D convolution of size $k$, where kernel size $k$ implies the coverage of local cross-channel interaction and the number of neighbors involved in attention prediction of one channel. To embody this process in equation:
\begin{equation}
    w_{eca} =\sigma(C1D_k(g(F_{FE}))),
\end{equation}
where $C1D_k$ denotes 1D convolution and $w_{eca}$ is the scale statistics of channel and $g(\cdot)$ stands for global average pooling. Then $F_{FE}$ is rescaled as
\begin{equation}
    \hat{F}_{FE}= w_{eca}\cdot F_{FE},
\end{equation}
where $\hat{F}_{FE}$ stands for rescaled feature. \\
\indent To utilize informative features from ECA, we apply residual block on the network. We transform residual block by applying  weighted summation on it
\begin{equation}
    F_{RECAB}= \frac{w_1}{\epsilon+\sum_{i=1}^{2}w_i} F_{input} + \frac{w_2}{\epsilon+\sum_{i=1}^{2}w_i} \hat{F}_{FE},
\end{equation}
where $F_{RECAB}$ is the final extracted feature and $w_i$ is a learnable weight which is a scalar per feature. As applying ReLU each $w_i$, we ensure $w_{i}\geq 0$, and fix $\epsilon$ value as 0.00001 to avoid numerical instability.
Similar to interpolation, the values of each weight are ranged from 0 to 1. Since these two weight values are learnable parameters, they find more proper values for producing well-balanced features in every training process. From the process, the informative feature is generated by considering the interdependencies among feature channels and thus it brings connectivity among channels and discriminative ability in network. \\
\textbf{Edge Profile(EP) module}: In SISR, it is significant point to maintain structure for high quality SR image.
\begin{figure}[h]
\begin{center}

\includegraphics[width=8.0cm]{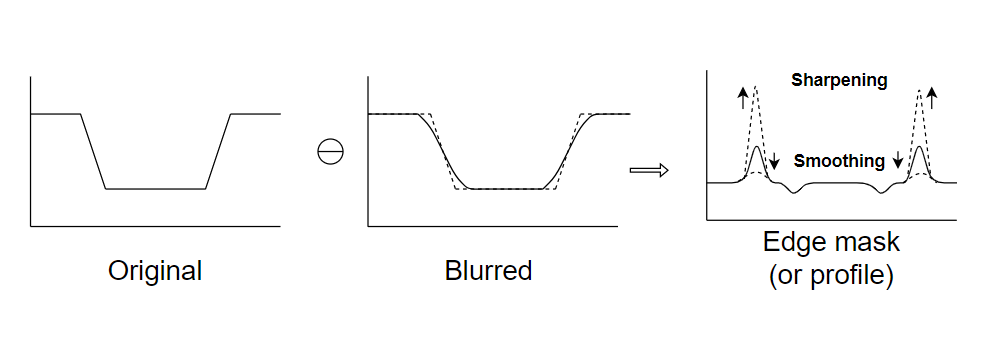}

\end{center}
   \caption{Extracting edge profile process}
\label{Fig.3}
\end{figure}
For considering structural information, we construct an EP module based on conventional image processing principle. This module extracts edge profile itself from the systemically organized feature by RECAB.  Intuitively, edge area has rapid variance of pixel as Fig.\ref{Fig.3}. This means that there are large pixel gradient values in edge area. Next, the onset and end of discontinuities (e.g. step and ramp discontinuities) in image are also described as edge areas. To extract edge profile of image, we consider utilizing discontinuous property of edge. As described in Fig.\ref{Fig.3}, to get edge mask(or profile), we subtract the blurred image from the original. So this process can be formulated as:
 \begin{equation}
    g(x,y) = f(x,y)- \Bar{f}(x,y),
    \label{Eq.14}
 \end{equation}
 where $g(x,y)$, $f(x,y)$ and $\Bar{f}(x,y)$ are edge mask, original image and blurred image respectively. We convert this process to deep learning method. First, we generate an image from feature $F_{RECAB}$, which comes from feature extraction, using one convolution layer:
 \begin{equation}
     I_{block SR}= H_{block SR}(F_{RECAB}),
 \end{equation}
 where $I_{block SR}$ is a produced image from feature $F_{RECAB}$, and $H_{block SR}(\cdot)$ can be denoted as image reconstruction in EP module, which generates RGB-channel image from the 64-channel feature.
 To form a blurred image, we transfer arithmetic mean filter concept using average pooling. Let's denote $S_{xy}$ as the set of coordinates in a rectangular sub-image window of size $m\times n$ where center point is $(x,y)$. Then this filter computes the average values of the original image $i(x,y)$ in the area defined by $S_{xy}$. In other words,  
 \begin{equation}
     \hat{i}(x,y) = \frac{1}{mn}\sum_{(s,t)\in S_{xy}}i(s,t),
 \end{equation}
where $\hat{i}(x,y)$ is a blurred image of $i(x,y)$. From this operation, if we define window size as $3\times3$, it can also be average pooling operation. So we form the blurred image by using it. 
\begin{equation}
    I_{blockblur} = H_{blur}(I_{block SR}),
\end{equation}
where $I_{block blur}$ is the blurred image from $I_{block SR}$ and $H_{blur }(\cdot)$ denotes average pooling whose kernel size is $3\times3$ and padding margin is $1$. From operation in Eq.\ref{Eq.14}, to get a edge profile(or mask), $I_{block SR}$ is subtracted by $I_{block blur}$. Then we apply ReLU operation on edge profile(or mask) for getting outer line. 
\begin{equation}
    M = ReLU(I_{block SR} \ominus I_{block blur}),
\end{equation}
where $M$ denotes edge profile(or mask) in REPB and  $\ominus$ is element-wise subtraction. To guide edge in training process, we concatenate $I_{block SR}$ with $M$
\begin{equation}
    I_{guided} = Concat(I_{block SR},M),
\end{equation}
where $I_{guided}$ and $Concat(\cdot)$ denote a guided image and concatenation operation respectively.
In the end, to generate the feature of EP module, we apply one convolution layer, and then give the feature $F_{FE}$ information by using residual structure.
\begin{equation}    
    F_{EP} =  F_{FE}+H_{EP}(I_{guided}),
\end{equation}
where $F_{EP}$ stands for the feature from EP module, which channel size is $64$, and $H_{EP}(\cdot)$ denotes edge profile module of REPB. By extracting structure information itself, we can obtain structure preserving effects.\\
\begin{figure*}[t]{\adjustbox{width = 12cm}{\Huge
\begin{tabular}{cccccccc}
\includegraphics{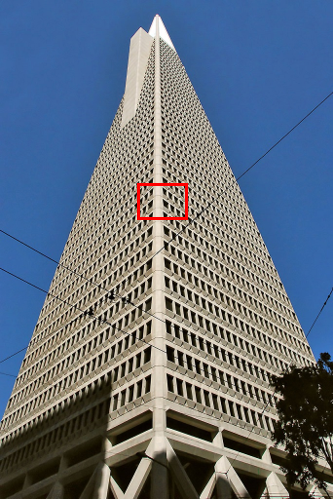}     & \includegraphics{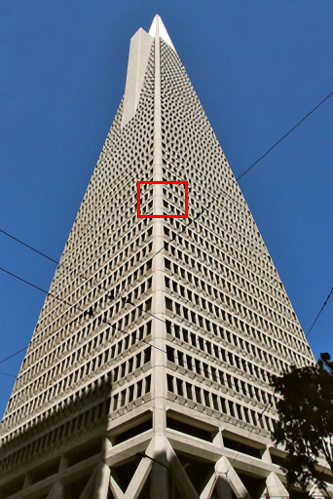}     & \includegraphics{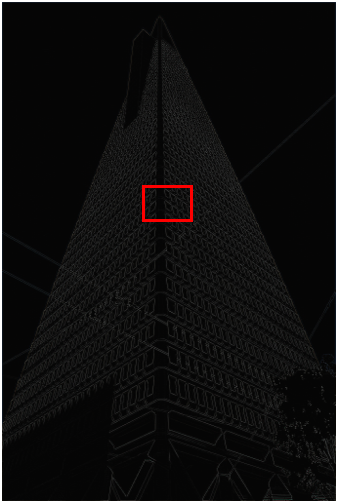}     & \includegraphics{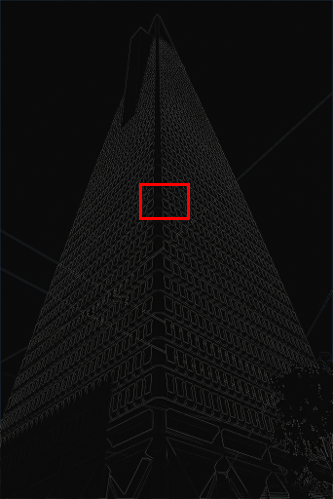}     & \includegraphics{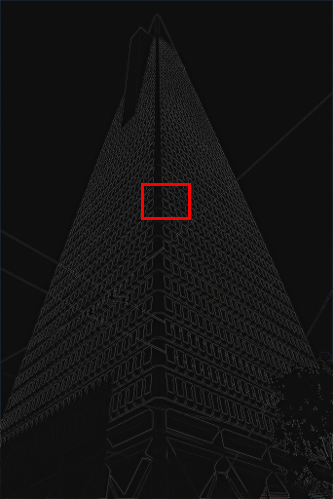}     & \includegraphics{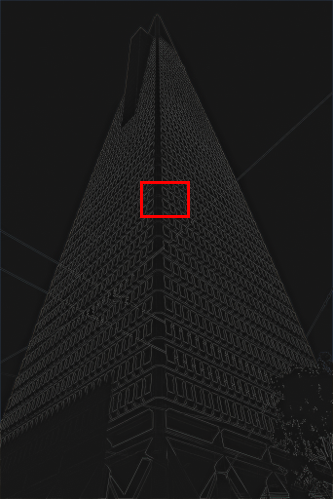}     & \includegraphics{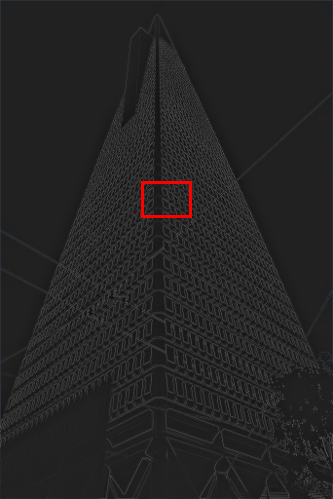}     & \includegraphics{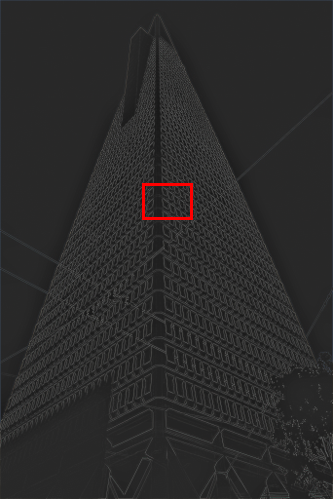}     \\
\includegraphics{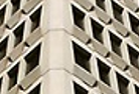} & \includegraphics{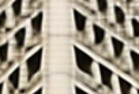} & \includegraphics{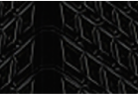} & \includegraphics{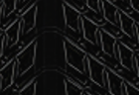} & \includegraphics{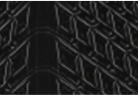} & \includegraphics{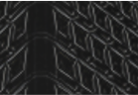} & \includegraphics{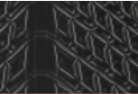} & \includegraphics{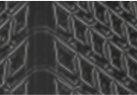} \\
(a) HR                                                     & (b) SR                                                     & (c) 24th Edge                                                 & (d) 48th Edge                                                 & (e) 72th Edge                                                 & (f) 96th Edge                                                 & (g) 112th Edge                                                & (h) Final edge                                               
\end{tabular}
}
}
\caption{Extracted edges of ``\,img\_048" by edge profile modules}
\label{Fig.5}
\end{figure*}
\textbf{Context Network(CN) module}: From EP module, we can obtain informative feature with structural information. This features could be beneficial to preserve structure. However, this module has limitation in handling high frequency components such as texture and sharpness of structure. Since the details have frequent pixel variations, it could be hard to capture. To reveal those contextual components, we construct CN module. \\
\indent Inspired by \cite{yu2015multi}, we design a CN module that is based on dilated convolutions. We apply CN module following EP module. As described in CN part of Fig.\ref{Fig.2}, our CN module consists of four $3\times3$ dilated convolution network, whose dilated factors are 1, 2, 4, and 1 in order. To prevent loss of resolution or coverage, we consider expansion of the receptive field to set up dilated factors exponentially. Intuitively, CN module can improve learning the feature maps by passing them through multiple layers that expose contextual information. After that, the output feature is added by the input feature as residual block.
\begin{equation}
{
    F_{CN} = F_{EP}
    +H_{f=1}\circ H_{f=4}\circ H_{f=2}\circ H_{f=1}(F_{EP}),
}
\end{equation}

where $F_{CN}$ is the output feature of CN module and $H_{f=n}(\cdot)$ denotes dilated convolution whose dilated factor $f$ is n. As this operation captures contextual information from the feature of EP module $F_{EP}$, our EPSR can minimize loss of texture and recover sharpness. In other words, recovering high frequency components can be ensured with minimizing side effects and damages.  \\
\subsection{Objective Functions}
Our EPSR is optimized with set-up loss functions. Normally, $L_{1}$\cite{lai2017deep,lai2018fast,lim2017enhanced,zhang2018residual}, $L_{2}$\cite{dong2015image,kim2016accurate,tai2017image,tai2017memnet}, adversarial and perceptual losses\cite{johnson2016perceptual,sajjadi2017enhancenet} have been used in SR method. To establish the effect of EPSR, we choose two loss functions $L_{1}$ and $L_{gradient}$. As proposed in previous works, we choose $L_{1}$ for guaranteeing stable convergence. Let's denote a given training set with $N$ LR images and their HR counterparts as $\{I^i_{LR},I^i_{HR}\}^N_{i=1}$, and then we can formulate  $L_1$ loss as:
\begin{equation}
  L_1 = \frac{1}{N}\sum_{n=1}^{N}||H_{EPSR}(I^i_{LR})-(I^i_{HR})||_{1}
\end{equation}
Since our EPSR utilizes diverse and different features each, REPBs generate edge profiles depending on feature information from feature input of them. To give consistent standard for EP modules in learning process, we consider loss function to guide them. By using Sobel filter\cite{parker2010algorithms}, we can extract gradient maps of HR and SR and formulate gradient loss function as:
\begin{equation}
L_{gradient}=\frac{1}{N}\sum_{n=1}^{N}||S(H_{EPSR}(I^i_{LR}))-S((I^i_{HR}))||_{1},
\end{equation}
where $S(\cdot)$ is gradient function based on Sobel filter\cite{parker2010algorithms}. By adding $L_{gradient}$ to $L_{1}$, we can achieve end-to-end network without additional module training. So the goal of training EPSR is to optimize the total loss function:
\begin{equation}
L_{Total}(\theta) = L_{1} +10^{-1} L_{gradient},
\end{equation}
where $\theta$ is the parameter set of EPSR. We set the coefficient as $10^{-1}$ empirically.
The loss function is optimized by ADAM gradient descent algorithm.
\section{Experiment Results}
\subsection{Settings}
 We state the settings of experiment about datasets, degradation models, evaluation, and training settings.\\ 
\textbf{Datasets.}
Following \cite{lim2017enhanced,zhang2018image,zhang2018residual}, we set up 800 high resolution images from DIV2K dataset \cite{timofte2017ntire} as a training set. For testing, we use 5 standard benchmark datasets: Set5\cite{bevilacqua2012low}, Set14\cite{zeyde2010single}, B100\cite{martin2001database}, Urban100\cite{huang2015single}, and Manga109\cite{matsui2017sketch}. \\
\textbf{Degradation Models.}
In order to prove the effectiveness of our EPSR, we use 3 degradation models to generate LR images. First, we generate LR images with scaling factor $\times2$, $\times3$, $\times4$ by using Bicubic Interpolation(BI) operation. Second, by using Gaussian kernel of size $7\times7$ with standard deviation 1.6, we blur HR image and downsample it with scaling factor $\times3$. We denote this process as BD\cite{zhang2018learning}.
At last, we downsample HR image with scaling factor $\times3$ using bicubic interpolation and then add Gaussian noise with level 30. This process is denoted as DN for short.\\
\textbf{Evaluation Metrics.}
The SR results are evaluated with PSNR and SSIM\cite{wang2004image} on Y channel(i.e. luminance) of YCbCr space.\\ 
\textbf{Training Settings.}
In training process, the training images are augmented by randomly rotating $90^{\circ}$,$180^{\circ}$,$270^{\circ}$, and horizontally flipping. In each training batch, 8 LR color patches with size $48\times48$ are extracted as input.
Our model is trained by ADAM optimizer with $\beta_1=0.9$, $\beta_2=0.99$, and $\epsilon=1e-8$. We set learning rate as $10^{-4}$ initially and then it is reduced to half every 200 epochs. We implement our proposed EPSR using Pytorch\cite{paszke2017and} on a Tesla V100 GPU.
\subsection{Ablation Study}
As we discussed above, our EPSR concentrates on structure preserving and representing details. To demonstrate effectiveness of our EPSR, we focus on showing influence of EP and CN modules, which could affect quality of SR results. Therefore, we set three comparisons by decomposing REPB, and two comparisons by feature extractions based on RECAB or RCAB by \cite{zhang2018image}.
\begin{table}[h]
\centering
\caption{Comparisons of models with different components. The best results are \textbf{highlighted.}}
{\adjustbox{width = 8cm}{
\begin{tabular}{c!{\vline[0.8pt]}c|c|c|c|c}
\Xhline{3\arrayrulewidth}
\multirow{2}{*}{} & \multicolumn{1}{c!{\vline[0.8pt]}}{Set5}      & \multicolumn{1}{c!{\vline[0.8pt]}}{Set14}     & \multicolumn{1}{c!{\vline[0.8pt]}}{BSD100}    & \multicolumn{1}{c!{\vline[0.8pt]}}{Urban100}  & Manga109     \\
                  & \multicolumn{1}{c!{\vline[0.8pt]}}{PSNR/SSIM} & \multicolumn{1}{c!{\vline[0.8pt]}}{PSNR/SSIM} & \multicolumn{1}{c!{\vline[0.8pt]}}{PSNR/SSIM} & \multicolumn{1}{c!{\vline[0.8pt]}}{PSNR/SSIM} & PSNR/SSIM    \\ \Xhline{3\arrayrulewidth}
RECAB               & 32.26/0.8937                   & 28.46/0.7802                    & 27.26/0.7327                   & 26.29/0.7934                   & 30.55/0.9017 \\ \Xhline{3\arrayrulewidth}
RECAB+EP            & 32.20/0.8932                   & 28.51/0.7823                   & 27.34/0.7356                   & 26.31/0.7943                   & 30.66/0.9062 \\ \Xhline{3\arrayrulewidth}
RECAB+CN            & 31.44/0.8786                   & 22.75/0.5959                   & 21.33/0.5206                   & 19.40/0.5848                   & 27.05/0.8274  \\ \hline\hline
RECAB+EP+CN         & \textbf{32.28}/\textbf{0.8945}                   & \textbf{28.55}/0.7828                   & \textbf{27.34}/\textbf{0.7362}                   & \textbf{26.43}/\textbf{0.7983}                   & \textbf{30.82}/\textbf{0.9084}
\\ \Xhline{3\arrayrulewidth}
RCAB+EP+CN          & 32.25/0.8939                   & 28.53/\textbf{0.7831}                   & 27.32/0.7349                   & 26.39/0.7975                   & 30.73/0.9076\\ \Xhline{3\arrayrulewidth}
\end{tabular}
}
}
\label{table.1}
\end{table}
First, to establish a criterion, we construct basic block without EP module and CN module. That is, by only RECAB, we generate SR images directly. As Fig.\ref{Fig.5}, only using RECAB is quite well in representing texture information. However, it has difficulty to recover image detail and edge components. Continually, we proceed with an experiment by connecting EP module to feature extraction for checking effect of edge profile. Even though edge profile is just provided on network, we can check enhancement of image reconstruction in aspect of structure preserving. Subsequently, adding CN at last, we build full REPB. As we explain details in section 3, CN helps to capture hidden information that include image details. In Fig.\ref{Fig.5} (e), which is generated by our EPSR, edge and texture information are reconstructed more stable than two images. In terms of PSNR and SSIM evaluations(See Table.\ref{table.1}) on all datasets, we can check that utilizing edge properties brings overall significant benefits in each evaluations. It implies that EP module is helpful to preserve structure image in reconstruction process as we can see in SSIM evaluations. Furthermore, by exploiting contextual information as image details such as texture and edge, CN module shows synergy effect with EP module. As a result, the efficacy of EP and CN modules is verified in images and numerical value evaluations both. Additionally, when we remove EP module in our EPSR, we can find some problems in recovering texture and edge information like as in Fig.\ref{Fig.5} (c). It shows that even if CN module gives benefits to capture contextual information, it could have weakness in exploiting overall features. we can also check this in numerical value evaluation. This indicates the rationality that CN module is plugged into the combination of RECAB and EP modules due to concentrating on capturing contextual information that contain image details not tendency of features.\\
\begin{figure}[b]{\adjustbox{width=12cm}
{\Huge
\begin{tabular}{cccccccccc}
\includegraphics{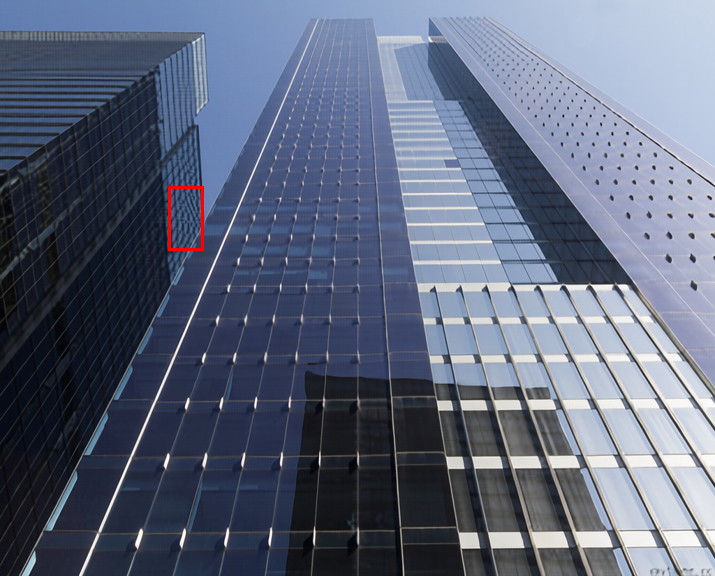} & \includegraphics{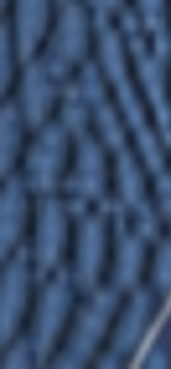} & \includegraphics{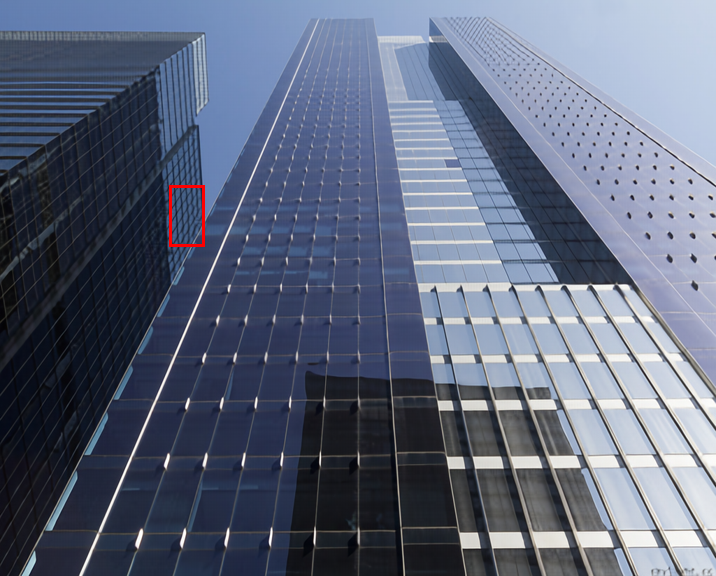} & \includegraphics{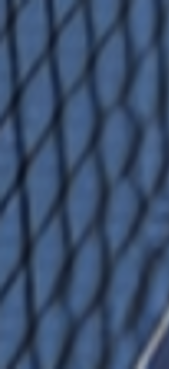} & \includegraphics{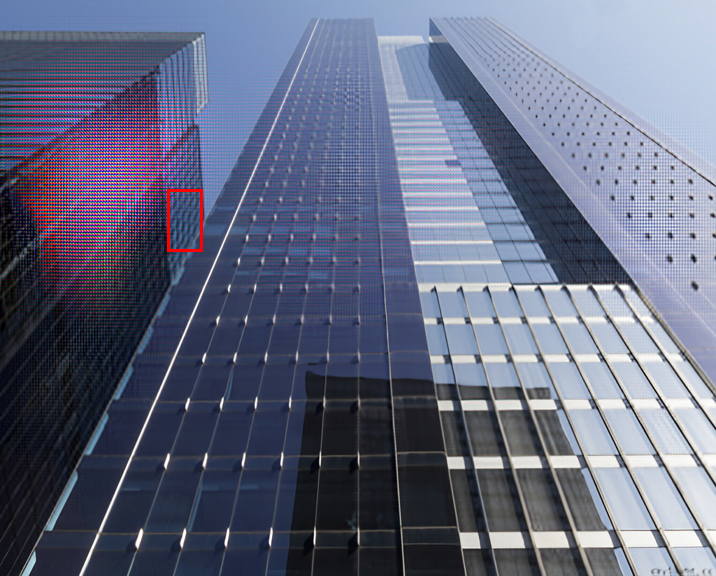} & \includegraphics{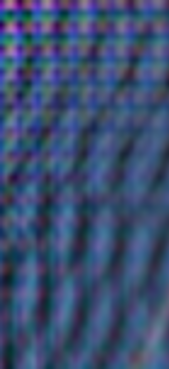} & \includegraphics{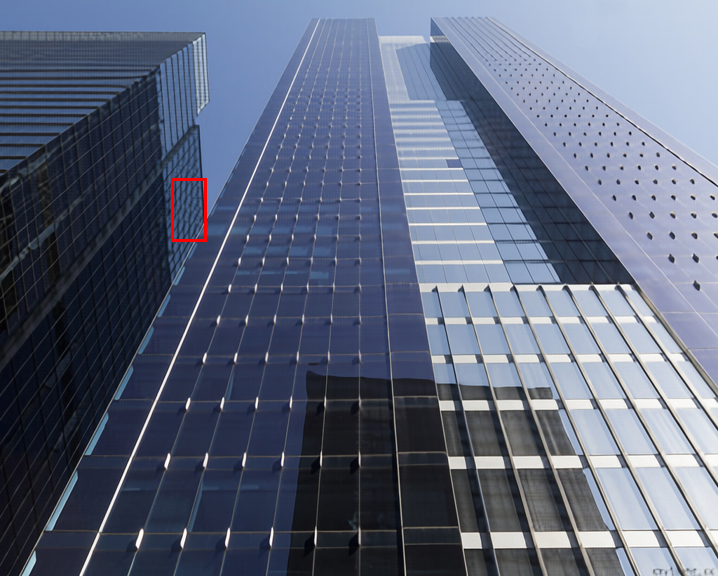} & \includegraphics{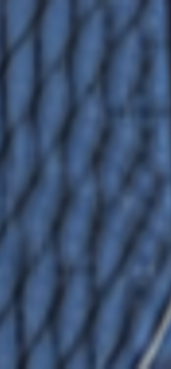} & \includegraphics{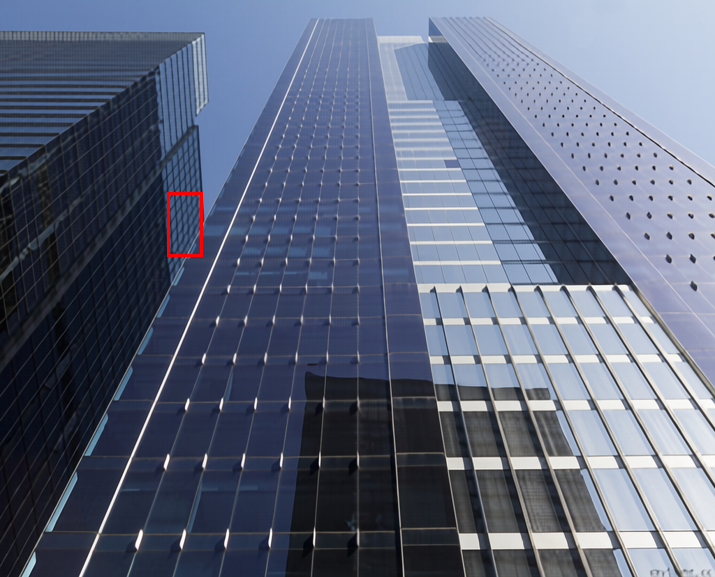} & \includegraphics{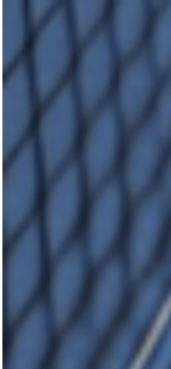} \\
\multicolumn{2}{c}{\fontsize{50}{60} \selectfont (a) RECAB}                                                                                           & \multicolumn{2}{c}{\fontsize{50}{60} \selectfont (b) RECAB + EP}                                                                                    & \multicolumn{2}{c}{\fontsize{50}{60} \selectfont (c) RECAB + CN}                                                                                      & \multicolumn{2}{c}{\fontsize{50}{60} \selectfont (d) RCAB+EP+CN}                                                                                 & \multicolumn{2}{c}{\fontsize{50}{60} \selectfont (e) RECAB+EP+CN}                                                                               
\end{tabular}
}
}
\caption{Ablation study with Bicubic(BI) degradation($\times4$) on ``\,img\_033" from Urban100.}
\label{Fig.5}
\end{figure}
\indent On top of that, we proceed extra experiments to investigate relationship between EP module and feature extraction. In our EPSR, we choose to use ECA for extracting features in RECAB. To verify the effect of it, we conduct experiment by substituting ECA for feature extraction to Channel Attention(CA) from \cite{zhang2018image}. we call the substitution as RCAB. As in Fig.\ref{Fig.5} (d), we can check that EPSR based on RCAB generates well SR image. However, we can see that the direction of edge lines are wrong way. Aforementioned in section 3, since CA has problem about channel predictions, it generates unclear features and it seems that EP module has some difficulties to find right edge lines. On the contrary, EPSR based on ECA feature extraction reconstructs edge and texture successfully. It is revealed visually that is generated by EPSR based on ECA feature extraction and in numerical value evaluations on PSNR and SSIM. This indicates that forming proper features is important key for extracting right edge profile to preserve structure in SR.

\begin{table}[h]
\centering
\caption{Quantitative results with BI degradation model. \textbf{Highlight} stands for the best performance, \textcolor{red}{red} indicates the second, and \textcolor{blue}{blue} is the third.}
{\adjustbox{width = 8cm}{
\begin{tabular}{|l|c|c|c|c|c|c|}
\hline
\multicolumn{1}{|c|}{\multirow{2}{*}{Method}} & \multirow{2}{*}{} & Set5         & Set14        & BSD100       & Urban100     & Manga109     \\ \cline{3-7} 
\multicolumn{1}{|c|}{}                        &                   & PSNR/SSIM    & PSNR/SSIM    & PSNR/SSIM    & PSNR/SSIM    & PSNR/SSIM    \\ \hline \hline
Bicubic                                       & 2                 & 33.66/0.9229 & 30.24/0.8688 & 29.56/0.8431 & 26.88/0.8403 & 30.80/0.9339 \\
SRCNN                                         & 2                 & 36.66/0.9542 & 32.45/0.9067 & 31.36/0.8879 & 29.50/0.8946 & 35.60/0.9663 \\
DEGREE                                        & 2                 & 37.40/0.9580 & 32.96/0.9115 & 31.73/0.8937 & -\quad  /\quad - & -\quad /\quad - \\
VDSR                                          & 2                 & 37.53/0.9587 & 33.05/0.9127 & 31.90/0.8960 & 30.77/0.9141 & 37.16/0.9740 \\
LapSRN                                        & 2                 & 37.52/0.9591 & 32.99/0.9124 & 31.80/0.8949 & 30.41/0.9101 & 37.53/0.9740 \\
EDSR                                          & 2                 & 37.99/0.9587 & 33.57/0.9175 & 32.16/0.8994 & 31.98/0.9272 & 39.10/0.9773 \\
MemNet                                        & 2                 & 37.78/0.9597 & 33.28/0.9142 & 32.08/0.8978 & 31.31/0.9195 & 37.72/0.9740 \\
IDN                                           & 2                 & 37.83/0.9600 & 33.30/0.9148 & 32.08/0.8985 & 31.27/0.9196 & 38.02/0.9749 \\
SRMDNF                                        & 2                 & 37.79/0.9601 & 33.32/0.9159 & 32.05/0.8985 & 31.33/0.9204 & 38.07/0.9761 \\
CARN                                          & 2                 & 37.76/0.9590 & 33.52/0.9166 & 32.09/0.8978 & 31.92/0.9256 & 38.36/0.9764 \\
RDN                                           & 2                 & \textcolor{blue}{38.24}/\textcolor{red}{0.9614} & \textcolor{blue}{34.01}/\textcolor{blue}{0.9212} & \textcolor{blue}{32.34}/\textcolor{blue}{0.9017} & \textcolor{blue}{32.89}/\textcolor{blue}{0.9353} & \textcolor{blue}{39.18}/\textcolor{blue}{0.9780} \\
RCAN                                          & 2                 & \textcolor{red}{38.27}/\textcolor{red}{0.9614} & \textcolor{red}{34.12}/\textcolor{red}{0.9216} & \textbf{32.41}/\textcolor{red}{0.9027} & \textcolor{red}{33.34}/\textcolor{red}{0.9384} & \textcolor{red}{39.44}/\textcolor{red}{0.9786} \\ \hline
EPSR                                         & 2                 & \textbf{38.29}/\textbf{0.9618} & \textbf{34.13}/\textbf{0.9227} & \textcolor{red}{32.38}/\textbf{0.9046} & \textbf{33.36}/\textbf{0.9401} & \textbf{39.57}/\textbf{0.9788} \\ \hline
\end{tabular}
}
}
\adjustbox{width= 8.0cm}{
\begin{tabular}{|l|c|c|c|c|c|c|}
\hline
Bicubic & 3 & 30.40/0.8686 & 27.54/0.7741 & 27.21/0.7389 & 24.46/0.7349 & 26.95/0.8556 \\
SRCNN   & 3 & 32.75/0.9090 & 29.29/0.8215 & 28.41/0.7863 & 26.24/0.7991 & 30.48/0.9117 \\
DEGREE  & 3 & 33.39/0.9182 & 29.61/0.8275 & 28.63/0.7921 & -\quad/\quad- & -\quad/\quad- \\
VDSR    & 3 & 33.66/0.9213 & 29.78/0.8318 & 28.83/0.7976 & 27.14/0.8279 & 32.01/0.9340 \\
LapSRN  & 3 & 33.82/0.9227 & 29.79/0.8320 & 28.82/0.7973 & 27.07/0.8271 & 32.21/0.9350 \\
EDSR    & 3 & 34.37/0.9270 & 30.28/0.8418 & 29.09/0.8052 & 28.15/0.8527 & 34.17/0.9476 \\
MemNet  & 3 & 34.09/0.9248 & 30.00/0.8350 & 28.96/0.8001 & 27.56/0.8376 & 32.51/0.9369 \\
IDN     & 3 & 34.11/0.9253 & 29.99/0.8354 & 28.95/0.8013 & 27.42/0.8359 & 32.69/0.9378 \\
SRMDNF  & 3 & 34.12/0.9254 & 30.04/0.8382 & 28.97/0.8025 & 27.57/0.8398 & 33.00/0.9403 \\
CARN    & 3 & 34.29/0.9255 & 30.29/0.8407 & 29.06/0.8034 & 28.06/0.8493 & 33.49/0.9440 \\
RDN     & 3 & \textcolor{blue}{34.71}/\textcolor{blue}{0.9296} & \textcolor{red}{30.57}/\textcolor{blue}{0.8468} & \textcolor{red}{29.26}/\textcolor{blue}{0.8093} & \textcolor{blue}{28.80}/\textcolor{red}{0.8653} & \textcolor{blue}{34.13}/\textcolor{blue}{0.9484} \\
RCAN    & 3 & \textbf{34.74}/\textbf{0.9299} & \textbf{30.65}/\textcolor{red}{0.8482} & \textbf{29.32}/\textcolor{red}{0.8111} & \textbf{29.09}/\textbf{0.8702} & \textcolor{red}{34.44}/\textbf{0.9499} \\ \hline
EPSR   & 3 & \textcolor{red}{34.73}/\textcolor{red}{0.9297} & \textcolor{blue}{30.52}/\textbf{0.8491} & \textcolor{blue}{29.15}/\textbf{0.8139} & \textcolor{red}{28.96}/\textbf{0.8702} & \textbf{34.46}/\textcolor{red}{0.9486} \\ \hline
\end{tabular}
}

{\adjustbox{width=8.0cm}{

\begin{tabular}{|l|c|c|c|c|c|c|}
\hline
Bicubic & 4 & 28.43/0.8109                                     & 26.00/0.7023                                     & 25.96/0.6678                                     & 23.14/0.6574                                     & \multicolumn{1}{c|}{25.15/0.7890}                                    \\
SRCNN   & 4 & 30.48/0.8628                                     & 27.50/0.7513                                     & 26.90/0.7103                                     & 24.52/0.7226                                     & \multicolumn{1}{c|}{27.66/0.8580}                                    \\
DEGREE  & 4 & 31.03/0.8761                                     & 27.73/0.7597                                     & 27.07/0.7177                                     & -\quad/\quad-                                    & \multicolumn{1}{c|}{-\quad/\quad-}                                   \\
VDSR    & 4 & 31.35/0.8838                                     & 28.02/0.7678                                     & 27.29/0.7252                                     & 25.18/0.7525                                     & \multicolumn{1}{c|}{28.82/0.8860}                                    \\
LapSRN  & 4 & 31.54/0.8866                                     & 28.09/0.7694                                     & 27.32/0.7264                                     & 25.21/0.7553                                     & \multicolumn{1}{c|}{29.09/0.8900}                                    \\
EDSR    & 4 & 32.09/0.8938                                     & 28.58/0.7813                                     & 27.57/0.7357                                     & 26.04/0.7849                                     & \multicolumn{1}{c|}{31.02/0.9148}                                    \\
MemNet  & 4 & 31.74/0.8893                                     & 28.26/0.7723                                     & 27.40/0.7281                                     & 25.50/0.7630                                     & \multicolumn{1}{c|}{29.42/0.8942}                                    \\
IDN     & 4 & 31.82/0.8903                                     & 28.25/0.7730                                     & 27.41/0.7297                                     & 25.41/0.7632                                     & 29.40/0.8936                                                         \\
SRMDNF  & 4 & 31.96/0.8925                                     & 28.35/0.7787                                     & 27.49/0.7337                                     & 25.68/0.7731                                     & \multicolumn{1}{c|}{30.09/0.9024}                                    \\
SRGAN   & 4 & 32.05/0.8910                                     & 28.53/0.7804                                     & 27.57/0.7354                                     & 26.07/0.7839                                     & \multicolumn{1}{c|}{-\quad/\quad-}                                   \\
NatSR   & 4 & 32.20/0.8939                                     & 28.54/0.7808                                     & 27.60/0.7366                                     & 26.21/0.7904                                     & \multicolumn{1}{c|}{-\quad/\quad-}                                   \\
SPSR    & 4 & 31.52/0.8827                                     & 27.74/0.7828                                     & 27.21/0.7276                                     & 24.80/0.8021                                     & \multicolumn{1}{c|}{30.12/0.9037}                                    \\
CARN    & 4 & 32.13/0.8937                                     & 28.60/0.7806                                     & 27.58/0.7349                                     & 26.07/0.7837                                     & \multicolumn{1}{c|}{30.40/0.9082}                                    \\
RDN     & 4 & \textcolor{red}{32.47}/\textcolor{red}{0.8990}   & \textcolor{red}{28.81}/\textcolor{red}{0.7871}   & \textcolor{red}{27.72}/\textcolor{red}{0.7419}   & \textcolor{blue}{26.61}/\textcolor{blue}{0.8028} & \multicolumn{1}{c|}{\textcolor{blue}{31.00}/\textcolor{red}{0.9151}} \\
RCAN    & 4 & \textbf{32.63}/\textbf{0.9002}                   & \textbf{28.87}/\textbf{0.7889}                   & \textbf{27.77}/\textbf{0.7436}                   & \textbf{26.82}/\textbf{0.8087}                   & \multicolumn{1}{c|}{\textbf{31.22}/\textbf{0.9173}}                  \\ \hline
EPSR    & 4 & \textcolor{blue}{32.42}/\textcolor{blue}{0.8969} & \textcolor{blue}{28.65}/\textcolor{blue}{0.7867} & \textcolor{blue}{27.45}/\textcolor{blue}{0.7403} & \textcolor{red}{26.64}/\textcolor{red}{0.8038}   & \multicolumn{1}{c|}{\textcolor{red}{31.16}/\textcolor{blue}{0.9127}} \\ \hline
\end{tabular}

}
}
\label{Table.2}
\end{table}

\subsection{Result with BI}
\textbf{Quantitative Comparison.}

To compare the effectiveness of our network with other methods, we investigate 14 state-of-the-art SR methods including general deep learning models, perceptual-driven models and edge-related models: SRCNN\cite{dong2015image}, DEGREE\cite{yang2017deep}, VDSR\cite{kim2016accurate}, LapSRN\cite{lai2017deep}, EDSR\cite{lim2017enhanced}, MemNet\cite{tai2017memnet}, IDN\cite{hui2018fast}, SRMDNF\cite{zhang2018learning}, CARN\cite{ahn2018fast}, RDN\cite{zhang2018residual}, RCAN\cite{zhang2018image}, SRGAN\cite{ledig2017photo}, NatSR\cite{soh2019natural}, SPSR\cite{ma2020structure}.
All of the quantitative comparisons for $\times2$, $\times3$ and $\times4$ SR are shown in Table.\ref{Table.2}. With rich texture information datasets, such as Set5, Set14, and BSD100, our EPSR obtains better results in SSIM compared to other networks. NatSR gets very high results, it shows weakness in BSD100 dataset specifically. However, our EPSR shows very well balanced results compared to NatSR and acquires high performance on all datasets. Furthermore, in PSNR, it obtains comparable results with RCAN and RDN whose main target is PSNR evaluation metric. In Urban100 and Manga109 datasets that contain rich repeated edge information, our EPSR achieves competitive results in PSNR and SSIM both. Subsequently, we compare our EPSR with SPSR and DEGREE which utilize structure information in super resolution method. They are dependent on artificial edge extracting work presents quite good improvement in structure preserving. However, the results do not reach on our EPSR. Overall, our EPSR shows high and competitive performance on PSNR and SSIM evaluation metrics.   \\
\textbf{Qualitative Comparison.}

\begin{figure}[h]
\centering
{\adjustbox{width=8.0cm}{\huge
\begin{tabular}{clccccc}
\multicolumn{2}{c}{\multirow{14}{*}{\includegraphics{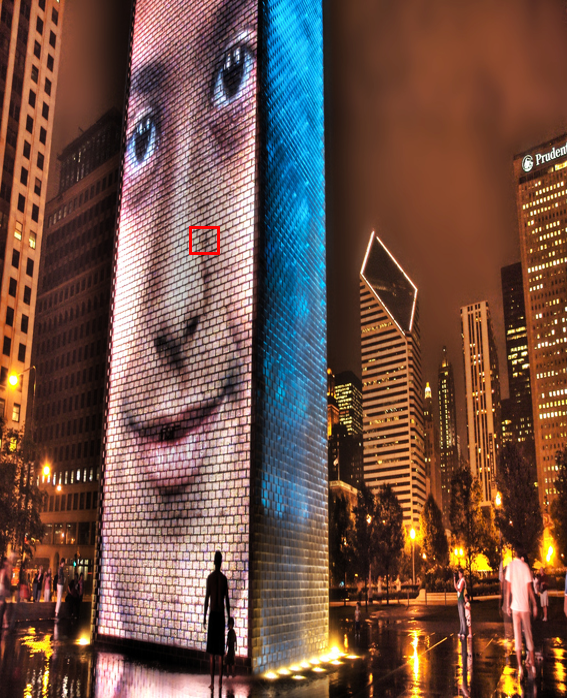}}} & \multirow{6}{*}{\includegraphics{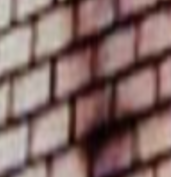}}   & \multirow{6}{*}{\includegraphics{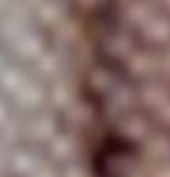}} & \multirow{6}{*}{\includegraphics{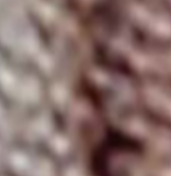}} & \multirow{6}{*}{\includegraphics{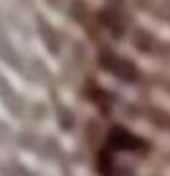}} & \multirow{6}{*}{\includegraphics{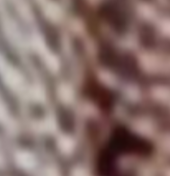}} \\
\multicolumn{2}{c}{}                                                    &                                                         &                                                            &                                                          &                                                           &                                                           \\
\multicolumn{2}{c}{}                                                    &                                                         &                                                            &                                                          &                                                           &                                                           \\
\multicolumn{2}{c}{}                                                    &                                                         &                                                            &                                                          &                                                           &                                                           \\
\multicolumn{2}{c}{}                                                    &                                                         &                                                            &                                                          &                                                           &                                                           \\
\multicolumn{2}{c}{}                                                    &                                                         &                                                            &                                                          &                                                           &                                                           \\
\multicolumn{2}{c}{}                                                    & HR                                                      & Bicubic                                                    & SRCNN\cite{dong2015image}                                                    & FSRCNN\cite{dong2016accelerating}                                                    & LapSRN\cite{lai2017deep}                                                    \\
\multicolumn{2}{c}{}                                                    & PSNR/SSIM                                               & 21.57/0.6281                                               & 22.03/0.6777                                             & 22.01/0.6772                                              & 22.01/0.6917                                              \\
\multicolumn{2}{c}{}                                                    & \multirow{6}{*}{\includegraphics{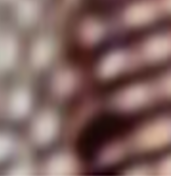}} & \multirow{6}{*}{\includegraphics{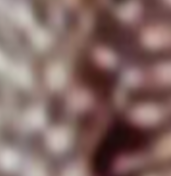}}  & \multirow{6}{*}{\includegraphics{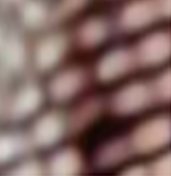}}   & \multirow{6}{*}{\includegraphics{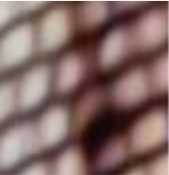}}   & \multirow{6}{*}{\includegraphics{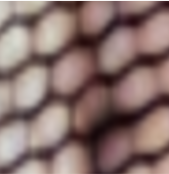}}   \\
\multicolumn{2}{c}{}                                                    &                                                         &                                                            &                                                          &                                                           &                                                           \\
\multicolumn{2}{c}{}                                                    &                                                         &                                                            &                                                          &                                                           &                                                           \\
\multicolumn{2}{c}{}                                                    &                                                         &                                                            &                                                          &                                                           &                                                           \\
\multicolumn{2}{c}{}                                                    &                                                         &                                                            &                                                          &                                                           &                                                           \\
\multicolumn{2}{c}{}                                                    &                                                         &                                                            &                                                          &                                                           &                                                           \\
\multicolumn{2}{c}{Urban100(x4):}                                       & EDSR\cite{lim2017enhanced}                                                       & SRMDNF\cite{zhang2018learning}                                                     & RDN\cite{zhang2018residual}                                                      & RCAN\cite{zhang2018image}                                                         & Ours                                                      \\
\multicolumn{2}{c}{img\_076}                                             & 23.94/0.7740                                            & 22.46/0.7108                                               & 24.08/0.7801                                             & \textcolor{red}{24.30}/0.7896                                              & 24.17/\textcolor{red}{0.7954}                                             
\end{tabular}
}}

{\adjustbox{width=8.0cm}{\huge
\begin{tabular}{clccccc}
\multicolumn{2}{c}{\multirow{14}{*}{\includegraphics{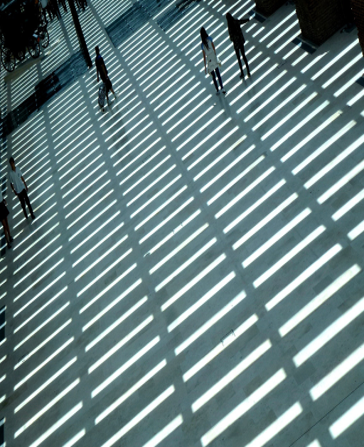}}} & \multirow{6}{*}{\includegraphics{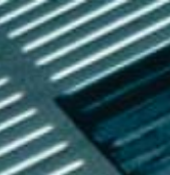}}     & \multirow{6}{*}{\includegraphics{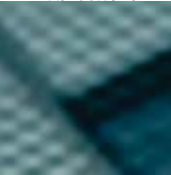}} & \multirow{6}{*}{\includegraphics{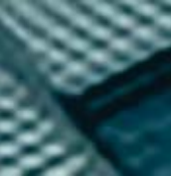}} & \multirow{6}{*}{\includegraphics{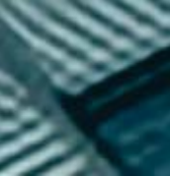}} & \multirow{6}{*}{\includegraphics{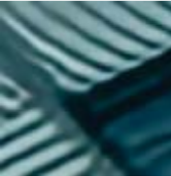}} \\
\multicolumn{2}{c}{}                                                    &                                                           &                                                            &                                                          &                                                           &                                                           \\
\multicolumn{2}{c}{}                                                    &                                                           &                                                            &                                                          &                                                           &                                                           \\
\multicolumn{2}{c}{}                                                    &                                                           &                                                            &                                                          &                                                           &                                                           \\
\multicolumn{2}{c}{}                                                    &                                                           &                                                            &                                                          &                                                           &                                                           \\
\multicolumn{2}{c}{}                                                    &                                                           &                                                            &                                                          &                                                           &                                                           \\
\multicolumn{2}{c}{}                                                    & HR                                                        & Bicubic                                                    & SRCNN\cite{dong2015image}                                                  & FSRCNN\cite{dong2016accelerating}                                                    & LapSRN\cite{lai2017deep}                                                    \\
\multicolumn{2}{c}{}                                                    & PSNR/SSIM                                                 & 23.62/0.8055                                               & 26.17/0.8757                                             & 26.68/0.8849                                              & 27.40/0.9136                                              \\
\multicolumn{2}{c}{}                                                    & \multirow{6}{*}{\includegraphics{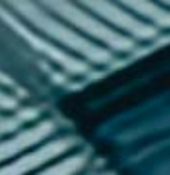}} & \multirow{6}{*}{\includegraphics{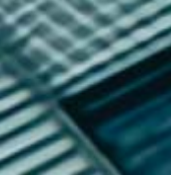}}    & \multirow{6}{*}{\includegraphics{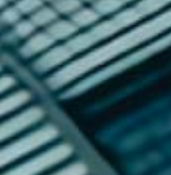}}   & \multirow{6}{*}{\includegraphics{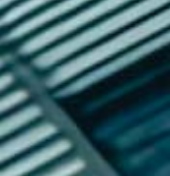}}   & \multirow{6}{*}{\includegraphics{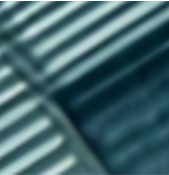}}   \\
\multicolumn{2}{c}{}                                                    &                                                           &                                                            &                                                          &                                                           &                                                           \\
\multicolumn{2}{c}{}                                                    &                                                           &                                                            &                                                          &                                                           &                                                           \\
\multicolumn{2}{c}{}                                                    &                                                           &                                                            &                                                          &                                                           &                                                           \\
\multicolumn{2}{c}{}                                                    &                                                           &                                                            &                                                          &                                                           &                                                           \\
\multicolumn{2}{c}{}                                                    &                                                           &                                                            &                                                          &                                                           &                                                           \\
\multicolumn{2}{c}{Urban100(x4):}                                       & SRMDNF\cite{zhang2018learning}                                                   & EDSR\cite{lim2017enhanced}                                                       & RDN\cite{zhang2018residual}                                                      & RCAN\cite{zhang2018image}                                                      & Ours                                                      \\
\multicolumn{2}{c}{img\_093}                                            & 28.13/0.9214                                              & 29.56/0.9326                                               & 28.16/0.9219                                             & 30.80/0.9420                                              & \textcolor{red}{31.68}/\textcolor{red}{0.9427}  
\end{tabular}
}}
{\adjustbox{width=8.0cm}{\huge

\begin{tabular}{clcccc}
\multicolumn{2}{c}{\multirow{14}{*}{\includegraphics{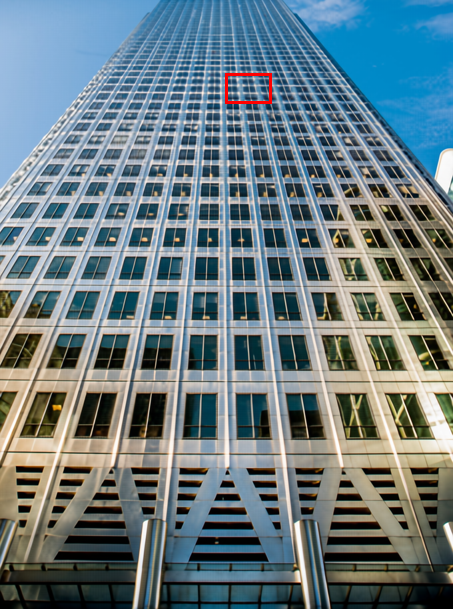}}} & \multirow{6}{*}{\includegraphics{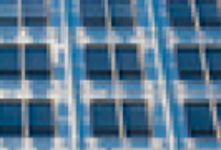}}    & \multirow{6}{*}{\includegraphics{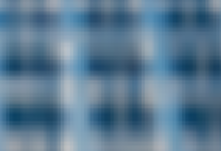}} & \multirow{6}{*}{\includegraphics{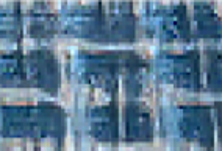}} & \multirow{6}{*}{\includegraphics{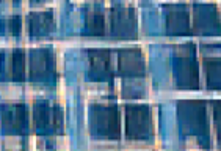}} \\
\multicolumn{2}{c}{}                                                    &                                                          &                                                            &                                                              &                                                           \\
\multicolumn{2}{c}{}                                                    &                                                          &                                                            &                                                              &                                                           \\
\multicolumn{2}{c}{}                                                    &                                                          &                                                            &                                                              &                                                           \\
\multicolumn{2}{c}{}                                                    &                                                          &                                                            &                                                              &                                                           \\
\multicolumn{2}{c}{}                                                    &                                                          &                                                            &                                                              &                                                           \\
\multicolumn{2}{c}{}                                                    & HR                                                       & Bicubic                                                    & EnhanceNet\cite{sajjadi2017enhancenet}                                                   & SFTGAN\cite{zhang2019sftgan}                                                    \\
\multicolumn{2}{c}{}                                                    & PSNR/SSIM                                                & 21.37/0.7133                                               & 23.76/0.7745                                                 & 24.34/0.8194                                              \\
\multicolumn{2}{c}{}                                                    & \multirow{6}{*}{\includegraphics{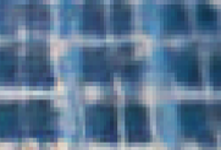}} & \multirow{6}{*}{\includegraphics{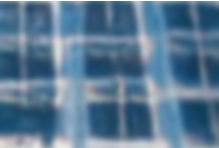}}   & \multirow{6}{*}{\includegraphics{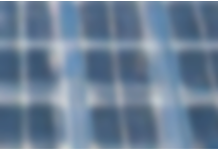}}      & \multirow{6}{*}{\includegraphics{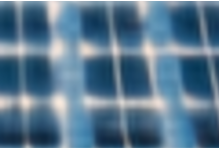}}   \\
\multicolumn{2}{c}{}                                                    &                                                          &                                                            &                                                              &                                                           \\
\multicolumn{2}{c}{}                                                    &                                                          &                                                            &                                                              &                                                           \\
\multicolumn{2}{c}{}                                                    &                                                          &                                                            &                                                              &                                                           \\
\multicolumn{2}{c}{}                                                    &                                                          &                                                            &                                                              &                                                           \\
\multicolumn{2}{c}{}                                                    &                                                          &                                                            &                                                              &                                                           \\
\multicolumn{2}{c}{Urban100(x4):}                                       & SRGAN\cite{wang2018esrgan}                                                    & NatSR\cite{soh2019natural}                                                      & SPSR\cite{ma2020structure}                                                         & Ours                                                      \\
\multicolumn{2}{c}{img\_030}                                            & 24.17/0.8212                                             & 25.23/0.8264                                               & 25.21/0.8398                                                 & \textcolor{red}{25.39}/\textcolor{red}{0.8426}                                             
\end{tabular}
}
}
\caption{Visual Comparison for SR($\times4$) with BI model on Urban100. The best results are \textcolor{red}{highlighted}}
\label{Figure.5}
\end{figure}
We present visual comparison on scale $\times4$. From Fig.\ref{Figure.5}, we see that our results are stronger in preserving structure and recovering texture both than other methods. 
  In ``\,img\_076\," and ``\,img\_093", we observe that most of the compared models cannot reconstruct the lattices and would have trouble in blurring effects.
 Other methods generate twisted lines and squashed the lattices. On the other hand, EPSR shows strength in recovering structural properties. We can see the capabilities of capturing structural characteristics of objects in image
and it contributes to preserving structure information in image and our EPSR captures image details well, which are including high frequency components.
In ``\,img\_030\," our EPSR shows clear structure in images without damage and distortion, while most of other methods fail to reconstruct fine appearance of the objects. The qualitative comparison verifies that our EPSR generate geometrically more stable image for perceptions by utilizing structural information extracted autonomously and exploiting contextual components.  \\

\subsection{Result with BD and DN}
\textbf{Quantitative Comparison.}

\begin{table}[h]
\centering
\caption{Quantitative results with BD degradation model}
{\adjustbox{width= 8.0cm}{
\begin{tabular}{|l|c|c|c|c|c|c|}
\hline
\multicolumn{1}{|c|}{\multirow{2}{*}{Method}} & \multirow{2}{*}{} & Set5         & Set14        & BSD100       & Urban100     & Manga109     \\ \cline{3-7} 
\multicolumn{1}{|c|}{}                        &                   & PSNR/SSIM    & PSNR/SSIM    & PSNR/SSIM    & PSNR/SSIM    & PSNR/SSIM    \\ \hline \hline
Bicubic                                       & 3                 & 28.78/0.8308 & 26.38/0.7271 & 26.33/0.6918 & 26.88/0.8403 & 25.46/0.8149 \\
SRMSR                                         & 3                 & 32.21/0.9001 & 28.89/0.8105 & 28.13/0.7740 & 25.84/0.7856 & 29.64/0.9003 \\
SRCNN                                         & 3                 & 32.05/0.8944 & 28.80/0.8074 & 28.13/0.7736 & 25.70/0.7770 & 29.47/0.8924 \\
FSRCNN                                        & 3                 & 26.23/0.8124 & 24.44/0.7106 & 24.86/0.6832 & 22.04/0.6745 & 23.04/0.7927 \\
VDSR                                          & 3                 & 33.25/0.9150 & 29.46/0.8244 & 28.57/0.7893 & 26.61/0.8136 & 31.06/0.9234 \\
IRCNN\_G                                      & 3                 & 33.38/0.9182 & 29.63/0.8281 & 28.65/0.7922 & 26.77/0.8154 & 31.15/0.9245 \\
IRCNN\_C                                      & 3                 & 33.17/0.9157 & 29.55/0.8271 & 28.49/0.7886 & 26.47/0.8081 & 31.13/0.9236 \\
SRMDNF                                        & 3                 & 34.01/0.9242 & 30.11/0.8364 & 28.98/0.8009 & 27.50/0.8370 & 32.97/0.9391 \\
RDN                                           & 3                 & \textcolor{blue}{34.58}/\textcolor{red}{0.9280} & \textcolor{blue}{30.53}/\textcolor{blue}{0.8447} & \textcolor{red}{29.23}/\textcolor{blue}{0.8079} & \textcolor{blue}{28.46}/\textcolor{blue}{0.8582} & \textcolor{blue}{33.97}/\textcolor{blue}{0.9465} \\
RCAN                                          & 3                 & \textbf{34.70}/\textbf{0.9288} & \textbf{30.63}/\textcolor{red}{0.8462} & \textbf{29.32}/\textcolor{red}{0.8093} & \textcolor{red}{28.81}/\textcolor{red}{0.8645} & \textcolor{red}{34.38}/\textbf{0.9483} \\ \hline
EPSR                                         & 3                 & \textcolor{red}{34.68}/\textbf{0.9288} & \textcolor{red}{30.56}/\textbf{0.8484} & \textcolor{blue}{29.14}/\textbf{0.8130} & \textbf{28.83}/\textbf{0.8667} & \textbf{34.51}/\textcolor{red}{0.9476} \\ \hline
\end{tabular}
}
}
\label{Table.3}
\end{table}
\vspace{-0.5cm}
\begin{table}[h]
\centering
\caption{Quantitative results with DN degradation model}

{\adjustbox{width= 8cm}{
\begin{tabular}{|l|c|c|c|c|c|c|}
\hline
\multicolumn{1}{|c|}{\multirow{2}{*}{Method}} & \multirow{2}{*}{} & Set5         & Set14        & BSD100       & Urban100     & Manga109     \\ \cline{3-7} 
\multicolumn{1}{|c|}{}                        &                   & PSNR/SSIM    & PSNR/SSIM    & PSNR/SSIM    & PSNR/SSIM    & PSNR/SSIM    \\ \hline\hline
Bicubic                                       & 3                 & 24.01/0.5369 & 22.87/0.4724 & 22.92/0.4449 & 21.63/0.4687 & 23.01/0.5381 \\
SRCNN                                         & 3                 & 25.01/0.6950 & 23.78/0.5898 & 23.76/0.5538 & 21.90/0.5737 & 23.75/0.7148 \\
FSRCNN                                        & 3                 & 24.18/0.6932 & 23.02/0.5856 & 23.41/0.5556 & 21.15/0.5682 & 22.39/0.7111 \\
VDSR                                          & 3                 & 25.20/0.7183 & 24.00/0.6112 & 24.00/0.5749 & 22.22/0.6096 & 24.20/0.7525 \\
IRCNN\_G                                      & 3                 & 25.70/0.7379 & 24.45/0.6305 & 24.28/0.5900 & 22.90/0.6429 & 24.88/0.7765 \\
IRCNN\_C                                      & 3                 & \textcolor{blue}{27.48}/\textcolor{blue}{0.7925} & \textcolor{blue}{25.92}/\textcolor{blue}{0.6932} & \textcolor{blue}{25.55}/\textcolor{blue}{0.6481} & \textcolor{blue}{23.93}/\textcolor{blue}{0.6950} & \textcolor{blue}{26.07}/\textcolor{blue}{0.8253} \\
RDN                                           & 3                 & \textcolor{red}{28.47}/\textcolor{red}{0.8151} & \textbf{26.60}/\textcolor{red}{0.7101} & \textbf{25.93}/\textcolor{red}{0.6573} & \textcolor{red}{24.92}/\textcolor{red}{0.7364} &\textcolor{red}{28.00}/\textcolor{red}{0.8591} \\ \hline
EPSR                                         & 3                 & \textbf{28.53}/\textbf{0.8142} & \textcolor{red}{26.57}/\textbf{0.7105} & \textcolor{red}{25.86}/\textbf{0.6588} & \textbf{25.16}/\textbf{0.7477} & \textbf{28.20}/\textbf{0.8634} \\ \hline
\end{tabular}
}
}
\label{table.4}
\end{table}
We apply our EPSR with BD degradation model, which is used recently in \cite{zhang2018image}, and following \cite{zhang2018residual}, we further compare various SR methods on image with DN degradation model. We compare our EPSR with 8-state-of-the-art SR methods with $\times3$ scaling factors: SRMSR\cite{peleg2014statistical}, SRCNN\cite{dong2015image}, FSRCNN\cite{dong2016accelerating}, VDSR\cite{kim2016accurate}, IRCNN\cite{zhang2017learning}, SRMDNF\cite{zhang2018learning}, RDN\cite{zhang2018residual}, and RCAN\cite{zhang2018image}.
In Table.\ref{Table.3} and Table.\ref{table.4}, all of the results are stated explicitly. We can observe that our EPSR shows higher performance compared to other methods. These results imply that our EPSR is very effective method for various types of degradation models. \\
\noindent\textbf{Qualitative Comparison.}
 \begin{figure}[t]
\centering
{\adjustbox{width=11.0cm}{\huge
\begin{tabular}{clccccc}
\multicolumn{2}{c}{\multirow{14}{*}{\includegraphics{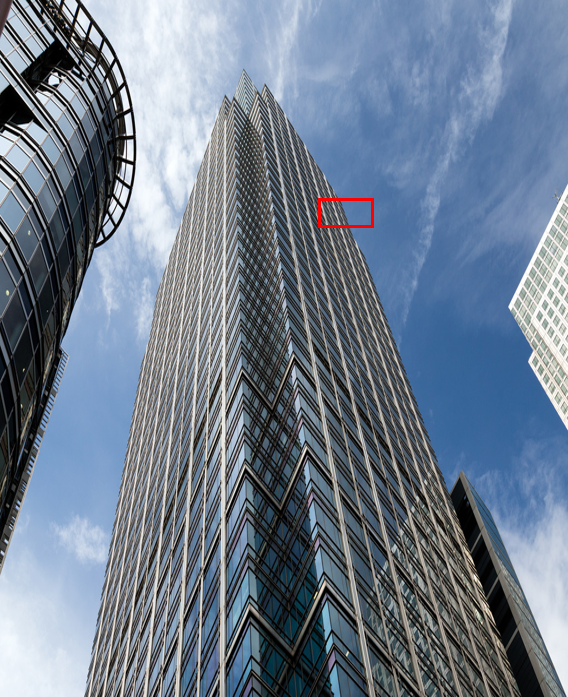}}} & \multirow{6}{*}{\includegraphics{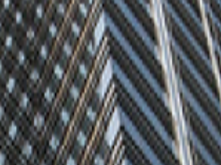}}      & \multirow{6}{*}{\includegraphics{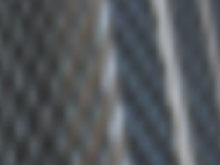}} & \multirow{6}{*}{\includegraphics{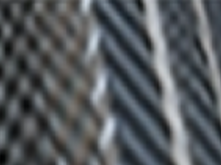}} & \multirow{6}{*}{\includegraphics{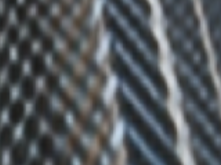}} & \multirow{6}{*}{\includegraphics{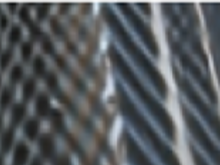}} \\
\multicolumn{2}{c}{}                                                    &                                                            &                                                            &                                                          &                                                          &                                                         \\
\multicolumn{2}{c}{}                                                    &                                                            &                                                            &                                                          &                                                          &                                                         \\
\multicolumn{2}{c}{}                                                    &                                                            &                                                            &                                                          &                                                          &                                                         \\
\multicolumn{2}{c}{}                                                    &                                                            &                                                            &                                                          &                                                          &                                                         \\
\multicolumn{2}{c}{}                                                    &                                                            &                                                            &                                                          &                                                          &                                                         \\
\multicolumn{2}{c}{}                                                    & HR                                                         & Bicubic                                                    & SRMSR\cite{peleg2014statistical}                                                    & SRCNN\cite{dong2015image}                                                    & VDSR\cite{kim2016accurate}                                                    \\
\multicolumn{2}{c}{}                                                    & PSNR/SSIM                                                  & 20.27/0.6480                                               & 22.05/0.7760                                             & 21.72/0.7566                                             & 22.32/0.7907                                            \\
\multicolumn{2}{c}{}                                                    & \multirow{6}{*}{\includegraphics{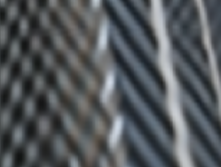}} & \multirow{6}{*}{\includegraphics{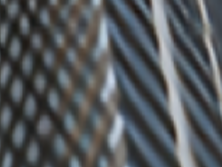}}  & \multirow{6}{*}{\includegraphics{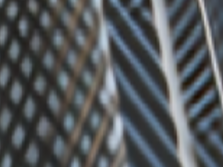}}   & \multirow{6}{*}{\includegraphics{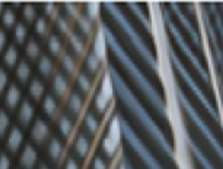}}  & \multirow{6}{*}{\includegraphics{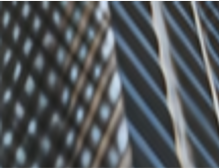}} \\
\multicolumn{2}{c}{}                                                    &                                                            &                                                            &                                                          &                                                          &                                                         \\
\multicolumn{2}{c}{}                                                    &                                                            &                                                            &                                                          &                                                          &                                                         \\
\multicolumn{2}{c}{}                                                    &                                                            &                                                            &                                                          &                                                          &                                                         \\
\multicolumn{2}{c}{}                                                    &                                                            &                                                            &                                                          &                                                          &                                                         \\
\multicolumn{2}{c}{}                                                    &                                                            &                                                            &                                                          &                                                          &                                                         \\
\multicolumn{2}{c}{Urban100(x4):}                                       & IRCNN\_G\cite{zhang2017learning}                                                    & SRMDNF\cite{zhang2018learning}                                                     & RDN\cite{zhang2018residual}                                                      & RCAN\cite{zhang2018image}                                                     & Ours                                                    \\
\multicolumn{2}{c}{img\_047}                                            & 22.48/0.7978                                               & 23.07/0.8269                                               & 23.74/0.8554                                             & 23.86/0.8623                                             & \textcolor{red}{24.16}/\textcolor{red}{0.8713}                                           
\end{tabular}
}}
\caption{Visual Comparison for SR($\times3$) with BD model on Urban100. The best results are \textcolor{red}{highlighted}.}
\label{Figure.7}
\end{figure}
We also show visual comparisons for challenging problem of blurring(BD) and noising(DN) degradation. First of all, in BD, there are difficulties in restoring definite texture and structural information. In Fig.\ref{Figure.7}, we can see that our results are clearer and more natural than other methods. Even though most methods suffer from heavy blurring problem, our EPSR recovers texture clearer than other methods. Especially, we can check the structure in our results are well-preserved without serious distortions. In succession, in DN, since there is heavy loss of information in LR, it is hard to reconstruct image ordinary. In Fig.\ref{Figure.8}, because of heavy damages of input, other methods have difficulties to overcome a lack of information and restore distortions. However, our EPSR is capable of restoring edge information with preventing texture loss.

This indicates that our EPSR can cope with damage and distortion of texture and structure by utilizing various information effectively. EPSR alleviates these troubles significantly and can reconstruct more details compared to other methods.

\begin{figure}[h]
\centering
{\adjustbox{width=11.0cm}{\huge

\begin{tabular}{cccccccc}
\includegraphics{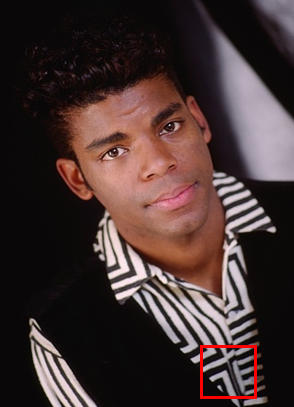} & \includegraphics{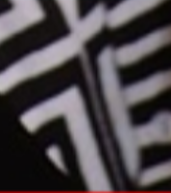} & \includegraphics{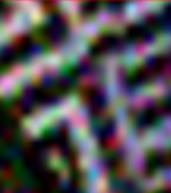} & \includegraphics{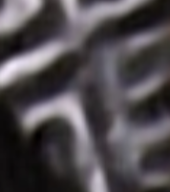} & \includegraphics{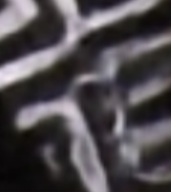} & \includegraphics{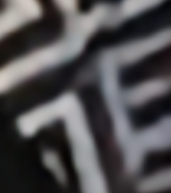} & \includegraphics{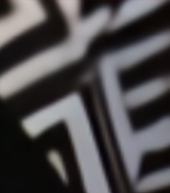} & \includegraphics{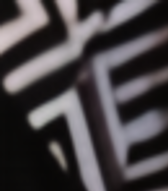} \\
B100(x3):                             & HR                                       & Bicubic                                       & SRCNN \cite{dong2015image}                                       & VDSR\cite{kim2016accurate}                                       & IRCNN\_C\cite{zhang2017learning}                                       & RDN\cite{zhang2018residual}                                       & Ours                                       \\
302008                                & PSNR/SSIM                                & 24.58/0.5737                                  & 25.60/0.8187                                & 25.77/0.8448                               & 28.45/0.8901                                  & 30.84/\textcolor{red}{0.9167}                              & \textcolor{red}{30.90}/0.9158                              
\end{tabular}

}
}
\caption{Visual Comparison for SR($\times3$) with DN model on Urban100. The best results are \textcolor{red}{highlighted}.}
\label{Figure.8}
\end{figure}

\section{Conclusion}
 In this paper, we propose Edge Profile Super Resolution (EPSR) method to preserve structure information and to restore texture in SISR. We construct EPSR by building modified-Fractal Residual Network (mFRN) structures hierarchically and repeatedly. mFRN is composed of residual Edge Profile Blocks (REPBs) consisting of three different modules such as Residual Efficient Channel Attention Block (RECAB) module, Edge Profile (EP) module, and Context Network (CN) module. RECAB generates more informative features with high frequency components. From the feature, EP module produce structure informed features by generating edge profile itself. Finally, CN module captures details by exploiting high frequency information such as texture and structure with proper sharpness. As repeating the procedure in mFRN structure, our EPSR could extract high-fidelity features and thus it prevents texture loss and preserves structure with appropriate sharpness. As our EPSR consider texture loss and structure information by applying conventional principle to deep learning method, high-quality results are obtained. Extension experiments on SR with BI, BD, and DN degradation models show the effectiveness of our EPSR.

%
%
%
 \bibliographystyle{splncs04}
 \bibliography{reference}
\end{document}